\documentstyle[aps,preprint]{revtex}
\newcommand{\be}{\begin{equation}}
\newcommand{\ee}{\end{equation}}
\newcommand{\beq}{\begin{eqnarray}}
\newcommand{\eeq}{\end{eqnarray}}
\newcommand{\mbf}{\mathbf}

\begin{document}
\draft
\tightenlines
\title{Stability of Synchronized Chaos in Coupled Dynamical Systems}
\author{Govindan Rangarajan$^{1,*}$ and
Mingzhou Ding$^{2,\dagger}$}
\address{$^1$Department of Mathematics and Center for Theoretical
Studies, Indian Institute of Science, Bangalore 560 012, India}
\address{$^2$Center for Complex Systems and Brain Sciences,
Florida Atlantic University, Boca Raton, FL 33431, USA}

\maketitle

\begin{abstract}
We consider the stability of synchronized chaos in coupled map
lattices and in coupled ordinary differential equations. Applying
the theory of Hermitian and positive semidefinite matrices we
prove two results that give simple bounds on coupling strengths
which ensure the stability of synchronized chaos. Previous results in
this area involving particular coupling schemes (e.g. global
coupling and nearest neighbor diffusive coupling) are included as
special cases of the present work.

\end{abstract}
\vskip 12pt \pacs{PACS Numbers: 05.40.+j, 05.45.+b}
\newpage

\section{Introduction}

Synchronization of coupled chaotic systems\cite{pecora1,pecora2}
has found applications in a variety of fields including
communications\cite{pecora1,cuomo}, optics\cite{winful,otsuko},
neural networks\cite{hansel1,hansel2,pasemann} and
geophysics\cite{vieira}. An essential prerequisite for these
applications is to know the bounds on the coupling strengths so
that the stability of the synchronous state is ensured. Previous
attempts\cite{heagy,ding,brown,glendinning,waller,afraimovich,rulkov,kaneko1,kaneko2,oppo,bunimovich,fujisaka,zhu,fink,belykh1,belykh2,zhan,nan,bressloff,hu}
aimed at obtaining such conditions have typically looked either at
systems of very small size or at very specific coupling schemes
(diffusive coupling, global all to all coupling etc. with a
single coupling strength). More recently, Pecora and
Carroll\cite{pecora3} introduced the notion of a master stability
function for general coupling topologies, but this function can
only be accessed in a numerical fashion. The contribution of this
Letter is a methodology that can lead to analytical bounds on the
individual coupling strengths under the mild assumption that the
coupling be symmetric. We demonstrate the method in the form of
two theorems, one applicable to coupled map lattices and the other
to coupled ordinary differential equations (ODEs).

\section{Stability Results}

The coupled map lattices we consider are in the form \be {\mbf
x}^i(n+1) = {\mbf f}({\mbf x}^i(n)) + \frac{1}{L} \sum_{j=1}^L
a_{ij} \left[ {\mbf f}({\mbf x}^j(n))-{\mbf f}({\mbf x}^i(n))
\right], \label{model}\ee and we use \be \dot{{\mbf x}}^i(t) =
{\mbf f}({\mbf x}^i(t)) + \sum_{j=1}^L b_{ij} {\mbf x}^j(t),
\label{model2}\ee for coupled ODEs. Here ${\mbf x}^i$ is an
$M$-dimensional state vector describing the $i$th map/ODE and
${\mbf f}$ is an $M$-dimensional map/vector field. We assume that
the coupling is symmetric, $a_{ij}=a_{ji}$ and $b_{ij}=b_{ji}$,
and that, in the absence of coupling, the individual
$M$-dimensional system is chaotic with the largest Lyapunov
exponent $h_{max}>0$. With the additional constraint that $
\sum_{j=1}^{L} b_{ij} = 0$ it is easy to see that the
synchronized state, ${\mbf x}^1 = {\mbf x}^2 = \cdots = {\mbf
x}^L = {\mbf x}$, is a solution to our models. We obtain
stability conditions by requiring that the ``transverse'' Lyapunov
exponents (which will be defined later) are all negative. We note
that the stability obtained here is in the sense of weak
stability (or Milnor stability), which ensures the transverse
stability of typical orbits, and is a necessary condition for
asymptotic stability where all orbits are transversely stable.
Our stability results for the synchronized chaotic solution are
expressed in the following two theorems.

{\bf Theorem 1}. For Eq.~(\ref{model}) the synchronized chaotic
state is stable if for all $i,j, \ j \neq i$, \be
[1-\exp(-h_{max})] < a_{ij} < [1+\exp(-h_{max})].
\label{stabf3}\ee

{\bf Theorem 2}. For Eq.~(\ref{model2}) the state of synchronized
chaos is stable if \be b_{ij}
> h_{max}/L, \ \ \ \forall \ i,j, \ j \neq i. \label{stabf4}\ee

We now sketch the proof of Theorem 1. Linearizing Eq.
(\ref{model}) around the synchronized chaotic state ${\mbf x}$ we
get : \be {\mbf z}^i(n+1) = {\mbf J}({\mbf x}(n)) \left[{\mbf
z}^i(n) (1- \frac{1}{L} \sum_{j \neq i} a_{ij}) + \frac{1}{L}
\sum_{j \neq i} a_{ij} {\mbf z}^j(n) \right], \label{linear}\ee
where ${\mbf z}^i$ denotes the $i$th map's deviations from ${\mbf
x}$ and ${\mbf J}$ is the $M \times M$ Jacobian matrix. In terms
of the $M \times L$ state matrix ${\mbf S}(n) = ({\mbf z}^1(n) \
{\mbf z}^2(n) \ \ldots \ {\mbf z}^L(n))$ Eq. (\ref{linear}) can be
written as the following matrix equation: \be {\mbf S}(n+1) =
{\mbf J}({\mbf x}(n)) {\mbf S}(n) {\mbf C}^T, \label{lin2}\ee
where ${\mbf C}^T$ is the transpose of the $L \times L$ coupling
matrix ${\mbf C}$ containing the coupling coefficients: \beq
[{\mbf C}]_{ii} & = & 1- \frac{1}{L} \sum_{j \neq i} a_{ij}, \ \
\ i=1,2, \ldots , L, \cr [{\mbf C}]_{ij} & = & a_{ij}/L, \ \ \ i
\neq j \label{C}\eeq

Since the coupling coefficients are assumed to be real
and symmetric, ${\mbf C}$
is Hermitian. It can be diagonalized and all
its eigenvalues are real\cite{horn}:
${\mbf C} = {\mbf E} {\mbf \Lambda} {\mbf E}^{-1}$.
Here ${\mbf E}$ and ${\mbf \Lambda}$ are the eigenvector
and eigenvalue matrices of ${\mbf C}$, respectively.
Let ${\mbf e}$ be one of the eigenvectors of ${\mbf E}$
and $\lambda$ its associated eigenvalue. Acting
Eq. (\ref{lin2}) on ${\mbf e}$ we get:
\be
{\mbf S}(n+1){\mbf e} = \lambda {\mbf J}({\mbf x}(n)) {\mbf S}(n) {\mbf e}.
\ee
Let ${\mbf u}(n) = {\mbf S}(n) {\mbf e}$. Then
\be
{\mbf u}(n+1) = \lambda {\mbf J}({\mbf x}(n)) {\mbf u}(n).
\label{linf}\ee

We now compute the Lyapunov exponents for the above reduced
system. We note that $\lambda=1$ is always an eigenvalue of
${\mbf C}$ and its corresponding eigenvector is $(1\ 1\ \ldots 1)^T$.
In this case, the above equation is just the linearization of
the individual map which was assumed to be chaotic. Therefore,
the eigenvector $(1\ 1\ \ldots 1)^T$
of ${\mbf C}$ with eigenvalue $1$ corresponds
to the synchronized chaotic state. The Lyapunov exponents in this
case are nothing but the Lyapunov exponents for the individual
system. Hence they are given by
$h_1=h_{max}$, $h_2$, ... , $h_M$. These describe the dynamics
within the synchronization manifold defined by
${\mbf x}^i = {\mbf x} \ \forall i$.

Next we consider the remaining eigenvalues and eigenvectors.
Since ${\mbf C}$ is a symmetric matrix, the remaining
eigenvectors span a $(L-1)$-dimensional subspace orthogonal
to the eigenvector $(1\ 1\ \ldots 1)^T$. Consequently,
this subspace is orthogonal to the synchronization manifold.
For each $\lambda \neq 1$ we calculate the Lyapunov
exponents for Eq. (\ref{linf}). Since $\lambda$ is a real
number (${\mbf C}$ being a symmetric matrix), the
Lyapunov exponents are easily calculated. Denoting them
by $\mu_1(\lambda)$, $\mu_2(\lambda)$, ..., $\mu_M(\lambda)$,
we have
\be
\mu_i(\lambda) = h_i + \ln |\lambda|, \ \ \ i=1,2, \ldots ,M.
\ee
We refer to these Lyapunov exponents as transversal
Lyapunov exponents\cite{ding} since they characterize the
behavior of infinitesimal vectors transversal to the
synchronization manifold. These determine the stability
of the synchronized chaotic state. For stability, we
require the transversal Lyapunov exponents for each
$\lambda \neq 1$ to be negative. This is equivalent to
the statement
\be
\mu_{max}(\lambda) = h_{max} + \ln |\lambda| < 0.
\ee
In other words, we require
$|\lambda| < \exp(-h_{max})$
for each $\lambda \neq 1$. From this equation
we see that for the synchronized chaotic state to be stable,
$\lambda=1$ should be the eigenvalue of ${\mbf C}$ with the
largest magnitude. Ordering the eigenvalues of ${\mbf C}$ as
$\lambda_1 = 1 > \lambda_2 \geq \lambda_3 \geq \cdots \geq \lambda_L$,
the stability conditions can be rewritten as
\beq
\label{stab2a}
\lambda_2 & < & \exp(-h_{max}), \\
\lambda_L & > & -\exp(-h_{max}). \label{stab2b}
\eeq

Our goal is to obtain bounds on $a_{ij}$ such that the above
two inequalities are simultaneously satisfied. Consider a
Hermitian matrix ${\mbf K}$ defined as follows:
\beq
[{\mbf K}]_{ii} & = & 1-\frac{(L-1)R}{L},
\ \ \ i=1,2, \ldots , L, \cr
[{\mbf K}]_{ij} & = & \frac{R}{L}, \ \ \ i \neq j,
\label{K}\eeq
where $R$ is a constant which will be characterized later.
Consider the matrix ${\mbf P}={\mbf K}-{\mbf C}$. We
see that the diagonal elements are given by
\be
{\mbf P}_{ii} = \frac{1}{L} \sum_{j \neq i} a_{ij} - \frac{(L-1)R}{L},
\ \ \ i=1,2, \ldots , L.
\label{Pdiag}\ee
These are positive if $a_{ij} > R \ \forall \ i,j, \ j \neq i$.
Next consider the absolute value of the off-diagonal elements
of ${\mbf P}$:
\be
|{\mbf P}_{ij}| = |\frac{R}{L} - \frac{a_{ij}}{L}|,
\ \ \ \forall \ i,j, \ j \neq i.
\ee
If $a_{ij} > R$, it can be seen that
\be
|{\mbf P}_{ii}| \geq \sum_{j \neq i} |{\mbf P}_{ij}|,
\ \ \ i=1,2, \ldots , L.
\ee
This implies that ${\mbf P}$ is positive semidefinite\cite{horn}.

We now introduce the concept of positive semidefinite
ordering\cite{horn}.
Since Hermitian matrices are generalizations of real numbers
and positive semidefinite matrices are generalizations of
nonnegative real numbers, one can introduce an ordering
among Hermitian matrices as follows\cite{horn}: Let
${\mbf A}$, ${\mbf B}$ be $L \times L$ Hermitian matrices.
We write ${\mbf A} \succeq {\mbf B}$ if the matrix ${\mbf A-B}$
is positive semidefinite. Further, if ${\mbf A} \succeq {\mbf B}$,
$\alpha_1 \geq \alpha_2 \geq \cdots \geq \alpha_L$ are the
ordered eigenvalues of ${\mbf A}$ and
$\beta_1 \geq \beta_2 \geq \cdots \geq \beta_L$ are the
ordered eigenvalues of ${\mbf B}$, then
\be
\alpha_i \geq \beta_i, \ \ \ i=1,2, \ldots , L.
\ee

Earlier, we had already shown that ${\mbf P} = {\mbf K}-{\mbf C}$
is positive semidefinite if $a_{ij} > R \ \forall \ i,j \ j \neq i$.
Since ${\mbf K}$ and ${\mbf C}$ are Hermitian, we have
${\mbf K} \succeq {\mbf C}$. Now the largest eigenvalues of
both ${\mbf K}$ and ${\mbf C}$ are equal to 1. The second
largest eigenvalue of ${\mbf K}$ can be easily calculated and
is equal to $1-R$. Therefore ${\mbf K} \succeq {\mbf C}$ implies
that
\be
\lambda_2 \leq (1-R).
\ee
Comparing with the inequality given in Eq. (\ref{stab2a}),
we see that this constraint is obeyed if $(1-R) < \exp(-h_{max})$.
That is,
\be
R > 1-\exp(-h_{max}).
\ee
But $\lambda_2 \leq 1-R$ only if $a_{ij} > R \ \forall \ i,j \ j \neq i$.
Putting the two inequalities together, we get the first stability
condition:
\be
a_{ij} > [1-\exp(-h_{max})], \ \ \ \forall \ i,j, \ j \neq i.
\label{stabf}\ee

Next we need to satisfy the second stability constraint given in
Eq. (\ref{stab2b}). Following a procedure similar to the one
above, we get the second stability condition: \be a_{ij} <
[1+\exp(-h_{max})], \ \ \ \forall \ i,j, \ j \neq i.
\label{stabf2}\ee Combining the stability conditions given in
Eqs. (\ref{stabf}) and (\ref{stabf2}) we get our final result as
follows. The synchronized chaotic state of Eq. (\ref{model}) is
stable (in the Milnor sense) if the coupling coefficients $a_{ij}$
(which are assumed to be symmetric) obey the stability condition
given in Eq. (\ref{stabf3}). Note that as $h_{max} \to 0$, the
range of allowed coupling strengths increases reaching a maximum
range of $(0,2)$. As $h_{max} \to \infty$, the range decreases to
zero. This confirms the intuitive expectation that it should be
harder to stabilize a synchronized state which is more chaotic.
The above result generalizes earlier
results\cite{ding,glendinning} obtained assuming that all
coupling coefficients are identical. Further, in some of the
earlier papers \cite{ding}, it was assumed that the constant
(dimensionless) coupling strength is less than 1 and hence the
upper bound given here (which is greater than 1) was not
explicitly observed in those previous studies.

We now numerically verify the above result by studying a system of
100 Henon maps. The $f$ [cf. Eq. (\ref{model})] for an individual
Henon map is given by:
\be
f_1(x_1,x_2) = 1+x_2-a x_1^2; \ \ \ f_2(x_1,x_2) = b x_2.
\ee
For the values $a=1.4$ and $b=0.3$, the maximum Lyapunov
exponent $h_{max}$ is found to be $0.43$. We now couple
100 Henon maps using the scheme given in Eq. (\ref{model})
where we randomly generate the coupling strengths $a_{ij}$'s.
From Eq. (\ref{stabf3}), the synchronized chaotic state is
stable if $0.35 < a_{ij} < 1.65$. We have numerically
verified this result.

If we have nearest neighbor coupling we can obviously obtain better
bounds than given in Eq. (\ref{stabf3}) since we know more
information about the coupling coefficients. In this case,
we obtain the following bounds (further details on this and
other coupling schemes can be found in our forthcoming
paper\cite{gr}):
\beq
\frac{1-\exp(-h_{max})}{1-\cos(2 \pi/L)} & < & a_{ij} <
\frac{1+\exp(-h_{max})}{1+\cos(2 \pi/L)},
\ \ \ {\rm if\ } L\ {\rm is\ odd}, \nonumber \\
\frac{1-\exp(-h_{max})}{1-\cos(2 \pi/L)} & < & a_{ij} <
\frac{1+\exp(-h_{max})}{2}, \ \ \ {\rm if\ } L\ {\rm is\ even}.
\nonumber \eeq Note that as $L$ becomes larger, the above range
shrinks to zero. We find\cite{gr} a conservative estimate for the
critical value $L_c$ which makes the range zero to be: \be L_c =
{\rm int} \left( \frac{2 \pi}{\cos^{-1}
\left[\exp(-h_{max})/(1+\exp(-h_{max})) \right]} \right). \ee
This generalizes the result found in Ref. \cite{ding}. The above
result implies that for a sufficiently large $L$ ($> L_c$) nearest
neighbor coupled systems can not have a stable synchronized
chaotic state.

We now prove Theorem 2. The initial treatment is similar to
the one used by Pecora and Carroll in arriving at the
master stability function\cite{pecora3}. The
structure of coupling that we have assumed includes the commonly
used diffusive coupling, nearest neighbor coupling, all-to-all
coupling, star coupling etc. Our proof breaks down if only one or
few of the components of ${\mbf x}^i$ are coupled. In this case,
Pecora and Carroll\cite{pecora3} have shown numerically that more
complicated stability conditions arise. Note that, unlike the
coupled map case, we only have a lower bound in the stability
condition. This difference arises from the fact that it is only
for maps the stability condition is in terms of the absolute
value of the eigenvalue which leads to both lower and upper
bounds. This is not so for coupled differential equations.

The proof of this theorem is along the same lines as
our proof for coupled maps. Linearizing around the synchronized
state we get
\be
\dot{{\mbf z}^i} = {\mbf J}({\mbf x}) {\mbf z}^i
+ \sum_{j \neq i} b_{ij} {\mbf z}^j,
\label{linear2}\ee
where ${\mbf z}^i$ denotes deviations from ${\mbf x}$ and
${\mbf J}$ is the $M \times M$ Jacobian matrix. We
now introduce the $M \times L$ state matrix ${\mbf S} =
({\mbf z}^1 \ {\mbf z}^2 \ \ldots \ {\mbf z}^L)$.
Then the linearized equation Eq. (\ref{linear2}) can be
written as the following matrix equation:
\be
\dot{{\mbf S}} = {\mbf J}({\mbf x}) {\mbf S} +  {\mbf S} {\mbf C}^T,
\label{lin4}\ee
where ${\mbf C}^T$ is the transpose of the $L \times L$
coupling matrix ${\mbf C}$ containing the
coupling coefficients:
\beq
[{\mbf C}]_{ii} & = & - \sum_{j \neq i} b_{ij},
\ \ \ i=1,2, \ldots , L, \cr
[{\mbf C}]_{ij} & = & b_{ij}, \ \ \ i \neq j
\label{C2}\eeq

Since the coupling is symmetric, ${\mbf C}$
is Hermitian. Following the same procedure as
before, we obtain
\be
\dot{{\mbf u}} = \left[ {\mbf J}({\mbf x})
+  \lambda {\mbf I} \right] {\mbf e},
\label{linf2}\ee
where ${\mbf I}$ is the $M \times M$ identity matrix,
${\mbf e}$ is an eigenvector of ${\mbf C}$ and
$\lambda$ its associated eigenvalue.
We now compute the Lyapunov exponents for the above reduced
system. We note that $\lambda=0$ is always an eigenvalue of
${\mbf C}$ and its corresponding eigenvector is $(1\ 1\ \ldots 1)^T$.
This corresponds
to the synchronized chaotic state. The Lyapunov exponents in this
case are nothing but the Lyapunov exponents for the individual
system. Hence they are given by
$h_1=h_{max}$, $h_2$, ... , $h_M$.

Next we consider the remaining eigenvalues and eigenvectors.
Since ${\mbf C}$ is a symmetric matrix, the remaining
eigenvectors span a $(L-1)$-dimensional subspace orthogonal
to the synchronization manifold.
For each $\lambda \neq 0$ we calculate the Lyapunov
exponents for Eq. (\ref{linf2}). Since $\lambda I$ commutes
with ${\mbf J}({\mbf x})$, the
Lyapunov exponents are easily calculated. Denoting them
by $\mu_1(\lambda)$, $\mu_2(\lambda)$, ..., $\mu_M(\lambda)$,
we have
\be
\mu_i(\lambda) = h_i + \lambda, \ \ \ i=1,2, \ldots ,M.
\ee
For stability, we
require these transversal Lyapunov exponents for each
$\lambda \neq 0$ to be negative. This is equivalent to
the statement
\be
\mu_{max}(\lambda) = h_{max} + \lambda < 0.
\ee
In other words, we require
$\lambda < -h_{max}$ for each $\lambda \neq 0$.
Ordering the eigenvalues of ${\mbf C}$ as
$\lambda_1 = 0 > \lambda_2 \geq \lambda_3 \geq \cdots \geq \lambda_L$,
the stability conditions can be rewritten as
\be
\label{stab6}
\lambda_2  <  -h_{max}.
\ee

As before, we wish to obtain bounds on $b_{ij}$ such that the
above inequality is satisfied. Consider a Hermitian matrix ${\mbf
K}'$ defined as follows: \beq [{\mbf K}']_{ii} & = & -(L-1)R', \
\ \ i=1,2, \ldots , L, \cr [{\mbf K}']_{ij} & = & R', \ \ \ i
\neq j, \label{Kprime}\eeq where $R'$ is a constant which will be
characterized later. Consider the matrix ${\mbf P}'={\mbf
K}'-{\mbf C}$. We see that the diagonal elements are given by \be
{\mbf P}'_{ii} = \sum_{j \neq i} b_{ij} - (L-1) R', \ \ \ i=1,2,
\ldots , L. \label{Pdiag2}\ee These are positive if $b_{ij} > R'
\ \forall \ i,j, \ j \neq i$. Next consider the absolute value of
the off-diagonal elements of ${\mbf P}'$: \be |{\mbf P}'_{ij}| =
|R' - b_{ij}|, \ \ \ \forall \ i,j, \ j \neq i. \ee If $b_{ij} >
R'$, it can be shown that \be |{\mbf P}'_{ii}| \geq \sum_{j \neq
i} |{\mbf P}'_{ij}|, \ \ \ i=1,2, \ldots , L \ee and ${\mbf
P}'_{ii} > 0$. This implies that ${\mbf P}'$ is positive
semidefinite\cite{horn}.

We have shown that ${\mbf P}' = {\mbf K}'-{\mbf C}$
is positive semidefinite if $b_{ij} > R' \ \forall \ i,j \ j \neq i$.
Since ${\mbf K}'$ and ${\mbf C}$ are Hermitian, we have
${\mbf K}' \succeq {\mbf C}$. Now the largest eigenvalues of
both ${\mbf K}'$ and ${\mbf C}$ are equal to 0. The second
largest eigenvalue of ${\mbf K}'$ can be easily calculated and
is equal to $-L R'$. Therefore ${\mbf K}' \succeq {\mbf C}$ implies
that
\be
\lambda_2 \leq - L R'.
\ee
Comparing with the inequality given in Eq. (\ref{stab6}),
we see that this constraint is obeyed if $- L R' < -h_{max}$.
That is,
\be
R' > h_{max}/L.
\ee
But $\lambda_2 \leq -L R'$ only if $b_{ij} > R' \ \forall \ i,j \ j \neq i$.
Putting the two inequalities together, we get the required stability
condition given in Eq. (\ref{stabf4}).

\section{Conclusions}

To conclude, we have derived very simple bounds on the
coupling coefficients which ensure the stability of the
synchronized chaotic state of $L$ symmetrically coupled
systems. We also gave specific bounds for the
nearest neighbor coupled map system. These results
allow for non equal coupling coefficients and generalize
earlier results found in the
literature\cite{ding,glendinning} which were obtained assuming that all
coupling coefficients are constant.
It is very easy to apply our
criteria to the system being studied and they encompasses a wide
class of coupling schemes including most of the popularly used
ones in the literature. Further, we expect the introduction
of non-equal coupling to lead to interesting new phenomena
in coupled systems. Our stability results would enable a
systematic exploration of such systems.

Our results were made possible by a sequence of operations. We
summarize them below since we feel that this approach is
applicable to the stability analysis of a wide variety of coupled
systems and not merely the specific problem considered in this
letter. First, we converted the linearized system to a matrix
equation. This was further simplified by looking at the evolution
of eigenmodes of the coupling matrix ${\mbf C}$. By realizing
that only the largest Lyapunov exponent matters for our analysis,
the stability conditions for the synchronized chaotic state were
recast as bounds on certain eigenvalues of the coupling matrix.
Then we bound the Hermitian matrix ${\mbf C}$ in the coupled map
system and the coupled oscillator system by carefully constructed
constant Hermitian matrices ${\mbf K}$ and ${\mbf K'}$
respectively. This was done in such a manner that that ${\mbf
K}-{\mbf C}$ and ${\mbf K'}-{\mbf C}$ are positive semidefinite
when the coupling coefficients satisfy certain inequalities.
Using a powerful result from matrix analysis, this automatically
implies that the eigenvalues of ${\mbf C}$ are bounded above by
the eigenvalues of ${\mbf K}$ (or ${\mbf K'}$). By comparing the
different bounds that we had derived, we then finally arrived at
Eqs. (\ref{stabf3}) and (\ref{stabf4}) which ensure the stability
of the synchronized chaotic state in the coupled map and coupled
oscillator systems respectively.

\section*{Acknowledgements}

The work was supported by US ONR Grant N00014-99-1-0062. GR thanks
Center for Complex Systems and Brain Sciences, Florida Atlantic
University, where this work was performed, for hospitality. He is
also associated with the Jawaharlal Nehru Center for Advanced
Scientific Research as a honorary faculty member and supported
by the Homi Bhabha Fellowship and DRDO and ISRO, India as part
of the Nonlinear Studies Group. We thank the referees for their
useful comments.

\end{document}